\documentclass[useAMS,usenatbib,usegraphicx]{mn2e}
\usepackage{bm}
\usepackage{times}

\newcommand{\Zsun}{\mathrm{Z}_{\sun}}
\newcommand{\Msun}{\mathrm{M}_{\sun}}
\newcommand{\nH}{n_{\mathrm{H}}}
\newcommand{\mH}{m_{\mathrm{H}}}
\newcommand{\kappap}{\kappa_{\mathrm{P}}}

\title[Dust coagulation in star formation]
{Dust coagulation in star formation with different metallicities}
\author[Hirashita \& Omukai]{Hiroyuki Hirashita$^1$\thanks{E-mail:
    hirashita@asiaa.sinica.edu.tw}
and Kazuyuki Omukai$^{2}$
\\
$^1$Institute of Astronomy and Astrophysics, Academia Sinica, P.O. Box
23-141, Taipei 10617, Taiwan\\
$^2$National Astronomical Observatory of Japan, Mitaka, Tokyo
181-8588, Japan
}
\date{2009 July 16}

\pagerange{\pageref{firstpage}--\pageref{lastpage}} \pubyear{2009}

\begin{document}
\label{firstpage}
\maketitle

\begin{abstract}
Dust grains coagulate into larger aggregates in dense gas.
This changes their size distribution and possibly affects
the thermal evolution of star-forming clouds. We here
investigate dust coagulation in collapsing
pre-stellar cores with different metallicities
by considering the thermal motions of grains.
We show that coagulation does occur even at low
metallicity $\sim 10^{-6}~\Zsun$. However, we also find
(i) that the H$_2$ formation rate on dust grains 
is reduced only after the majority of H$_2$ is formed; and
(ii) that the dust opacity is modified only after the core
becomes optically thick.
Therefore, we conclude that the effects of 
dust coagulation can safely be neglected in discussing 
the temperature evolution of the pre-stellar cores
for any metallicity as long as the grain motions are
thermal.
\end{abstract}


\section{Introduction}
The first generation of stars is believed to
be very massive 
($\ga 100~\Msun$; e.g.\ \citealt{bromm04}), 
while those in the solar neighbourhood have typical
masses of $\la\Msun$ \citep[e.g.][]{larson05}.
The transition in the characteristic stellar mass 
should have occurred in the history of the Universe.
The most widely accepted cause for this transition is metal enrichment 
in the interstellar medium \citep{omukai00,bromm01,schneider02}.
In other words, the transition of the characteristic
stellar mass occurs at a `critical metallicity'.

Dust grains should be produced in supernovae of first stars 
and thus be present in metal-enriched gas 
\citep{todini01,nozawa03,schneider04}. Their thermal
emission becomes effective for gas cooling through collisional
coupling at such high  density that the Jeans mass is
below $\Msun$. 
With a sufficient amount of dust, dust cooling is able 
to induce fragmentation into subsolar mass cores. 
If most of the metals condense into dust, 
the critical metallicity is as low as
$\sim 10^{-5}~\Zsun$ \citep{omukai05}, although
difference in grain composition and grain size distribution 
alters the exact value by
about an order of magnitude \citep{schneider06}.

In such dense clumps where dust cooling is 
efficient, however, coagulation of dust grains could proceed 
and modify the dust opacity as well as the H$_2$ formation 
efficiency on dust grains, consequently affecting the thermal
evolution of star-forming clouds. Therefore, coagulation may
have a large influence on the critical metallicity.
In spite of this potential importance, the effects of
coagulation on the thermal evolution
of star-forming clouds have not been
considered in previous studies.

In this paper, we assess the effects of coagulation on the
thermal evolution of star-forming clouds. We focus on
low metallicity in the context of the critical metallicity,
but we treat all the range of metallicity from nearly
zero to $\Zsun$. Therefore, the methods and results in this
paper are applicable to star-forming clouds at any
metallicity.
This paper is organized as follows. The method used in this
paper is explained in Section \ref{sec:method}. We describe
some basic properties of the thermal evolution of collapsing
cores in Section \ref{sec:without}. The effects
of coagulation are discussed in Section \ref{sec:effect}.
Finally, Section \ref{sec:conclusion} gives the conclusion.

\section{Method}\label{sec:method}

We examine the effect of coagulation on the thermal
evolution of star-forming collapsing cores. First, the
density and temperature evolution of collapsing cores
are calculated with a fixed grain size distribution
(i.e.\ without coagulation). Then, under these
calculated density and temperature, we estimate the
effect of dust coagulation. We explain the basic ingredients
of our calculations in the following with particular
emphasis on physics related to dust grains.

\subsection{Basic equations for pre-stellar
collapse}\label{subsec:review}

We calculate the thermal evolution of star-forming
cores by using the formulations developed by
\citet{omukai00} and improved by \citet{omukai05}.
Here we briefly review the methods.

The collapsing gas in a star-forming core is treated
as one zone with a homogeneous hydrogen
nuclei number density $\nH$ and a single gas
temperature $T$. These density and
temperature can be interpreted as those in the central
region of the core, whose size is given by the Jeans length. 
The evolution of density is calculated by
\begin{eqnarray}
\frac{\mathrm{d}\rho}{\mathrm{d}t}=
\frac{\rho}{t_\mathrm{ff}}~~~\mbox{with}~~~
t_\mathrm{ff}=\sqrt{\frac{3\pi}{32G\rho}},
\end{eqnarray}
where $t_\mathrm{ff}$ is the free-fall time and $G$ is
the gravitational constant. 
The temperature evolution is
calculated by solving the energy equation,
\begin{eqnarray}
\frac{\mathrm{d}e}{\mathrm{d}t}=-p
\frac{\mathrm{d}}{\mathrm{d}t}\left(\frac{1}{\rho}
\right) -\Lambda_\mathrm{net},
\end{eqnarray}
where $e$ is the specific thermal energy, $p$ is the
pressure, $\rho$ is the density, and
$\Lambda_\mathrm{net}$ is the net cooling rate.
The net cooling rate includes all the important processes:
molecular/atomic line emission, dust and gas continuum emission, 
and chemical heating and cooling.
For the details of the cooling/heating
processes and of their treatment, we refer to
\citet{omukai00}, although we review the processes
related to dust grains in the following subsections.
We additionally impose a cosmic background radiation at
$1+z=10$ (i.e.\ $T_{\rm CMB}= 27$ K), so
that the temperature does not drop below it. However,
our results on coagulation are not sensitive to the
background temperature.

\subsection{Dust cooling}

Dust grains achieve the following balance between
dust thermal emission and heating due to collisions
with gas particles:
\begin{eqnarray}
4\sigma (T_\mathrm{gr}^4-T_\mathrm{CMB}^4) \kappap\beta_\mathrm{esc}
\rho\mathcal{D}=H_\mathrm{gr},
\end{eqnarray}
where $\sigma$ is the Stefan--Boltzmann constant,
$T_\mathrm{gr}$ is the dust temperature, $\kappap$ is
the Planck mean opacity of dust grains per unit dust
mass, $\beta_\mathrm{esc}$ is the
photon escape probability, $\mathcal{D}$ is the
dust-to-gas mass ratio, and $H_\mathrm{gr}$ is the
collisional heating (cooling) rate of dust (gas)
particles. Following \citet{omukai00}, the
photon escape probability is written as
$\beta_\mathrm{esc}=\min (1,\,\tau^{-2})$ by using 
the optical depth $\tau$.

The collisional heating rate of dust particles can
be written as \citep{hollenbach79}
\begin{eqnarray}
H_\mathrm{gr}=
\frac{n_\mathrm{gr}(2k_\mathrm{B}T
-2k_\mathrm{B}T_\mathrm{gr})}{t_\mathrm{coll}},
\end{eqnarray}
where
$t_\mathrm{coll}^{-1}=\nH\sigma_\mathrm{gr}
\bar{v}_\mathrm{H}f$ is the mean interval between two
successive collisions, $n_\mathrm{gr}$ and
$\sigma_\mathrm{gr}$ are the grain number density and
cross-section, respectively,
$\bar{v}_\mathrm{H}=({8k_\mathrm{B}T}/{\pi\mH})^{1/2}$
is the average speed of hydrogen nuclei under a
Maxwellian distribution, and $f=0.3536+0.5y_\mathrm{He}$
($y_\mathrm{He}=0.083$, corresponding to a He mass fraction
of $Y=0.25$) measures the contribution from species
other than protons.
Once the grain temperature exceeds
the sublimation temperature, $T_\mathrm{sub}$, the grains
are assumed to disappear.

The grain cross-section is related to the total surface
area of grains per unit dust mass, $S$, as
\begin{eqnarray}
n_\mathrm{gr}\sigma_\mathrm{gr}=\nH\mH
(1+4y_\mathrm{He})S\mathcal{D}.\label{eq:crosssection}
\end{eqnarray}
Given the grain size distribution functions per unit
dust mass, $\mathcal{N}_{\mathrm{gr},i}(a)$ (i.e.\
$\mathcal{N}_{\mathrm{gr},i}(a)\,\mathrm{d}a$ gives the
number of dust
grains of species $i$ with radius between $a$ and
$a+\mathrm{d}a$ per unit dust mass
and the grains are assumed to be spherical),
and the mass fractional abundance, $f_i$, of each grain
species $i$, $S$ is quantified as
\begin{eqnarray}
S=\sum_if_iS_i~~~\mbox{with}~~~S_i=\int_0^\infty
\mathcal{N}_{\mathrm{gr},i}(a)\pi a^2\,\mathrm{d}a.
\end{eqnarray}
The grain surface area determines the H$_2$ formation
rate on grain surface, and the line emission from
formed H$_2$ contributes to gas cooling.

The total absorption coefficient at frequency $\nu$ 
per unit dust mass is written in a similar way as
\begin{eqnarray}
\kappa (\nu )=\sum_if_i\kappa_{\nu ,i}~~\mbox{with}~~
\kappa_{\nu ,i}=\int_0^\infty Q_{\nu}^i(a)
\mathcal{N}_{\mathrm{gr},i}(a)\pi a^2\mathrm{d}a,\label{eq:kappa_nu}
\end{eqnarray}
where $Q_{\nu}^i(a)$ is the absorption cross-section
of species $i$ normalized to the geometrical cross-section.

\subsection{H$_2$ formation on dust grains}

The H$_2$ formation rate on grain surface per unit volume
can be expressed as
$R_\mathrm{H_2} 
=k_\mathrm{gr}n(\mathrm{H})\nH$,
where $n(\mathrm{H})$ is the number density of H atoms in
the gas phase.
We use the rate coefficient
\citep{tielens85}
\begin{eqnarray}
k_\mathrm{gr}=6\times 10^{-17}(S/S_{\sun})
\sqrt{T/300~\mathrm{K}}\, f_\mathrm{a}S_\mathrm{H},
\label{eq:rate}
\end{eqnarray}
where
$S$ is the total surface area of dust grains, 
$S_{\sun}$ is that for the MRN grain size
distribution
(Section \ref{subsec:parameter}) for $Z=\Zsun$
($Z$ is the metallicity),  
\begin{equation}
f_\mathrm{a}=\left\{1+\exp\left[7.5 \times 10^{2}
(1/75-1/T_{\rm gr})\right]\right\}^{-1},
\end{equation}
and
\begin{eqnarray}
S_\mathrm{H}=[1+0.04(T+T_\mathrm{gr})^{0.5}
+0.002T+8\times 10^{-6}T^2]^{-1}.
\end{eqnarray}
Note that, without coagulation, $S/S_{\sun}$ is simply
$Z/\Zsun$. The rate coefficient
(equation \ref{eq:rate}) with $S/S_{\sun}=1$ is
consistent with observationally derived
rates for a sample of clouds in the solar neighbourhood
\citep{jura75,hollenbach79}. Thus we
expect that equation (\ref{eq:rate}) with $S/S_{\sun}=1$
is valid for the MRN
grain size distribution and solar metallicity.

\subsection{Grain coagulation}\label{subsec:coag}

The time evolution of the grain size distribution
by coagulation is calculated by adopting the formulation
of \citet{hirashita09}. Since they treated coagulation
under a constant density, we modified the formulation
to treat the varying gas density in collapsing cores.
We briefly review the calculation
method. The details are described in \citet{hirashita09}.
In this subsection, we omit the index $i$
specifying the species.

As assumed in \citet{hirashita09}, we
consider grains to be compact and spherical. In other
words, we do not
consider fluffy aggregates, which are suggested to
form after sticking of grains
\citep{meakin88,ossenkopf93}.\footnote{We note that
aggregation rather than coagulation may be more
appropriate terminology to express the grain growth after
grain--grain collisions, especially when we consider
fluffy aggregates. However, we use the term
`coagulation' throughout this paper, since we only
consider compact (i.e.\ not fluffy) grains.}
There are two competing effects of the
fluffiness. Fluffy grains would increase the
coagulation rate
\citep{ormel07}, which could
enhance the decrease of $S$. On the other hand,
fluffy grains have larger $S$ than compact grains.
Because of these competing effects on $S$,
it is not obvious if the change of $S$
is under- or overestimated by assuming a compact structure.

We solve the coagulation equation discretized for the
grain size.
The total grain mass density is normalized to
$\nH\mH (1+4y_\mathrm{He})\mathcal{D}$.
Thus, as $\nH$ becomes denser in the evolution of
collapsing cores, the grain density
also becomes proportionally larger.


We consider silicate and graphite as grain species. In
order to avoid complexity in compound species, we only
treat collisions between the same species.
The time evolution of $\tilde{\rho}_k$ (grain density
contained in the $k$-th size bin; all the subscripts
in the following equation specify the bin number) by
coagulation can be written as
\begin{eqnarray}
\frac{\mathrm{d}\tilde{\rho}_k}{\mathrm{d}t}=-m_k
\tilde{\rho}_k
\sum_{\ell =1}^{N}\alpha_{\ell k}\tilde{\rho}_\ell+
\frac{1}{2}\sum_{j=1}^{N}\sum_{\ell =1}^N\alpha_{\ell j}
\tilde{\rho}_\ell
\tilde{\rho}_jm_\mathrm{coag}^{\ell j}(k),
\end{eqnarray}
where
\begin{eqnarray}
\alpha_{\ell k}=\left\{
\begin{array}{ll}
{\displaystyle
\frac{\beta\sigma_{\ell k}v_{\ell k}}{m_km_\ell}
} &
\mbox{if $v_{\ell k}<v_\mathrm{coag}^{\ell k}$,} \\
0 & \mbox{otherwise,}
\end{array}
\right.
\end{eqnarray}
($\beta$ is the sticking probability) and
$m_\mathrm{coag}^{\ell j}(k)=m_k$ if the mass of the
coagulated particle $m_\ell +m_j$ is in the mass range
of the $k$-th bin;
otherwise $m_\mathrm{coag}^{\ell j}(k)=0$. Coagulation
is assumed to occur only if the relative velocity is less
than the coagulation
threshold velocity $v_\mathrm{coag}^{\ell k}$.
The cross-section for the coagulation is
$\sigma_{\ell k}=\pi (a_\ell +a_k)^2$.
The coagulation
threshold velocity is given by
\citep*{chokshi93,dominik97,yan04}
\begin{eqnarray}
v_\mathrm{coag}^{\ell k}=21.4\left[
\frac{a_\ell^3+a_k^3}{(a_\ell +a_k)^3}\right]^{1/2}
\frac{\gamma^{5/6}}{E^{1/3}R_{\ell k}^{5/6}
\rho_\mathrm{gr}^{1/2}},\label{eq:vcoag}
\end{eqnarray}
where $\gamma$ is the surface energy per unit area,
$R_{\ell k}\equiv a_\ell a_k/(a_\ell +a_k)$ is the reduced
radius of the
grains, $E$ is related to Poisson's ratios ($\nu_\ell$ and
$\nu_k$) and Young's modulus ($E_\ell$ and $E_k$) by
$1/E\equiv (1-\nu_\ell )^2/E_\ell +(1-\nu_k)^2/E_k$. 
We also assume $\beta =1$ for the sticking probability.
A sticking probability of order unity below the coagulation
threshold is shown by experimental studies \citep{blum00}.
The effect of $\beta$ is also addressed in
Section \ref{subsec:surface}.

We adopt a single velocity for each grain size. The velocity
is given by the thermal (Brownian) velocity as
\begin{eqnarray}
v_k=\left(\frac{8k_\mathrm{B}T}{\pi m_k}\right)^{1/2}.
\end{eqnarray}
Smaller grains have larger thermal velocities, and
the effect of coagulation first appears for the smallest
grains (see also Section \ref{sec:effect}). As shown in
\citet{ossenkopf93}, motions other than the thermal
motion are potentially important. In particular, large
grains tend to be coupled with turbulent motion if
it exists
\citep{ossenkopf93,weidenschilling94,ormel09}.\footnote{Recently,
\citet{ormel09} show that even if turbulent velocity is
considered, large grains with $a\ga 0.1~\mu$m do not
coagulate significantly on the free-fall timescale.
However, they consider relatively small densities
($\la 10^7$ cm$^{-3}$), and the effect of turbulence on the
higher density range is still open.}
For the sake of simplicity and exploratory nature of this
work, the contribution of other velocity sources is
ignored here and remains open for further study.

Each time-step is divided into four equal small steps, and we
apply $v_{k\ell}=v_k+v_\ell$, $|v_k-v_\ell|$, $v_k$, and
$v_\ell$ in each step, following \citet{hirashita09}.
A relative velocity of
$v_{k\ell}=({8k_\mathrm{B}T}/{\pi\mu})^{1/2}$, where
$\mu$ the reduced mass, is often adopted, but the difference
caused by the different treatment of relative velocity
is too small to change our conclusion.

\subsection{Adopted parameters for dust grains}
\label{subsec:parameter}

As stated above, we assume a mixture of silicate and
graphite species. We adopt $\mathcal{D}=0.006(Z/\Zsun )$
for both silicate and graphite, and the contribution
from these two species are summed with weights
$f_i=0.54$ and 0.46 for silicate and graphite,
respectively \citep{hirashita09}.
The dust-to-gas ratio in solar metallicity environments
(0.006) is taken from \citet{spitzer78}.
The material densities $s$ are adopted from \citet{draine84}
(3.3 and 2.26 g cm$^{-3}$ for silicate and graphite,
respectively).


The quantities for coagulation
are taken from \citet{chokshi93}
(the data for
quartz and graphite are used for silicate and graphite,
respectively; i.e.\ $\gamma =25$ and 12~erg~cm$^{-2}$,
$E=5.4\times 10^{11}$ and $3.4\times 10^{10}$ dyn cm$^{-2}$,
and $\nu =0.17$ and 0.5 for silicate and graphite,
respectively).
The sublimation temperatures ($T_\mathrm{sub}=1400$ and
1750 K for
silicate and graphite, respectively) are taken from
\citet{laor93}.

The absorption cross-section of grains is calculated
by using the Mie theory \citep{bohren83}. The optical
constants of silicate (astronomical silicate) and
graphite are taken from \citet{draine84}.
The grain size distribution is assumed to be
$\mathcal{N}_{\mathrm{gr},i}(a)\propto a^{-3.5}$ in
the size range
$0.005~\mu\mathrm{m}\leq a\leq 0.25~\mu$m
and $\mathcal{N}_{\mathrm{gr},i}(a)=0$ out of this range
for both silicate
and graphite as indicated by \citet{mathis77} (MRN).

\section{Thermal evolution without coagulation}
\label{sec:without}

First, density and temperature evolutions are calculated
without coagulation (i.e.\ with the fixed MRN
grain size distribution; Section \ref{subsec:parameter}).
Then, the density and temperature given in
this section are used as the background density and
temperature under which grain coagulation is
calculated in the next section.



\begin{figure}
\includegraphics[width=0.45\textwidth]{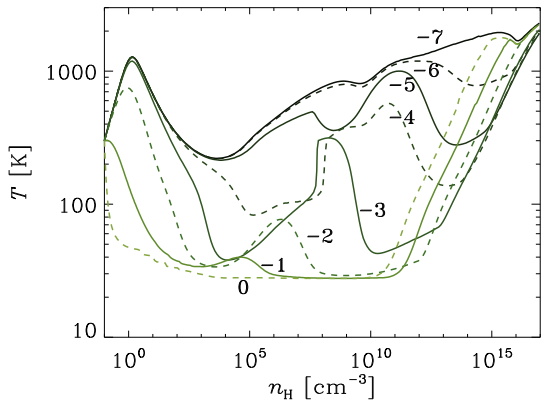}
 \caption{Temperature evolution of collapsing pre-stellar cores 
with different metallicities as a function of the hydrogen 
number density, which increases with time. 
Solid curves indicate the cases with metallicities of
$\log (Z/\Zsun )=-7$, $-5$, $-3$, and $-1$, while dashed
curves show the results with
$\log (Z/\Zsun )=-6$, $-4$, $-2$, and 0. The numbers
represent the values of $\log (Z/\Zsun )$ adopted
for individual lines.}
 \label{fig:nT}
\end{figure}

The evolutions of collapsing cores are shown in
Fig.\ \ref{fig:nT} in the metallicity range
$10^{-7}\le Z/\Zsun\le 1$.
In the case with $10^{-7}\Zsun$, metal effects are negligible 
and its temperature evolution is identical to the metal-free one.  
In cores with metallicity $\la 10^{-4}~\Zsun$, 
H$_2$ is mainly formed via gas-phase reactions.
With higher metallicity, i.e.\ higher dust abundance, the
majority of H$_2$ is formed via the dust surface reactions.
H$_2$ line cooling contributes significantly 
to the thermal evolution at
relatively low densities
($\nH\la 10^{5\mbox{--}8}$~cm$^{-3}$) as seen by the
dips of the temperature evolution in this density range,
although carbon and oxygen fine-structure
lines contribute more to the cooling
at $Z\ga 10^{-2}~\Zsun$. 
Therefore, the modification of grain surface area (if any) 
is the most relevant around this density range. 
On the other hand, the impact of
dust cooling appears around
$\nH\sim 10^{10\mbox{--}13}$ cm$^{-3}$ for
metallicities $Z\ga 10^{-6}\Zsun$ as seen by the
dips of the temperature evolution. 
At higher densities, dust cooling has minor influence on
the thermal evolution, since the system becomes
optically thick to dust continuum.
Although the metallicity range where the effect of dust cooling 
appears is broadly consistent with \citet{omukai05}, 
the effect is more pronounced at the same metallicity 
in our case, especially for the lowest values $\sim 10^{-6}\Zsun$.
This is because our assumed composition of carbonaceous grains, 
graphite, is more refractory than organic compounds adopted
in \citet{omukai05}. Grains survive up to
$\nH\sim 10^{15}$ cm$^{-3}$ in our case.
Our choice is conservative in the sense that the coagulation 
effect becomes more visible as the dust survives and coagulates
up to higher temperature and density.

\section{Effects of coagulation}\label{sec:effect}

\begin{figure*}
\includegraphics[width=0.45\textwidth]{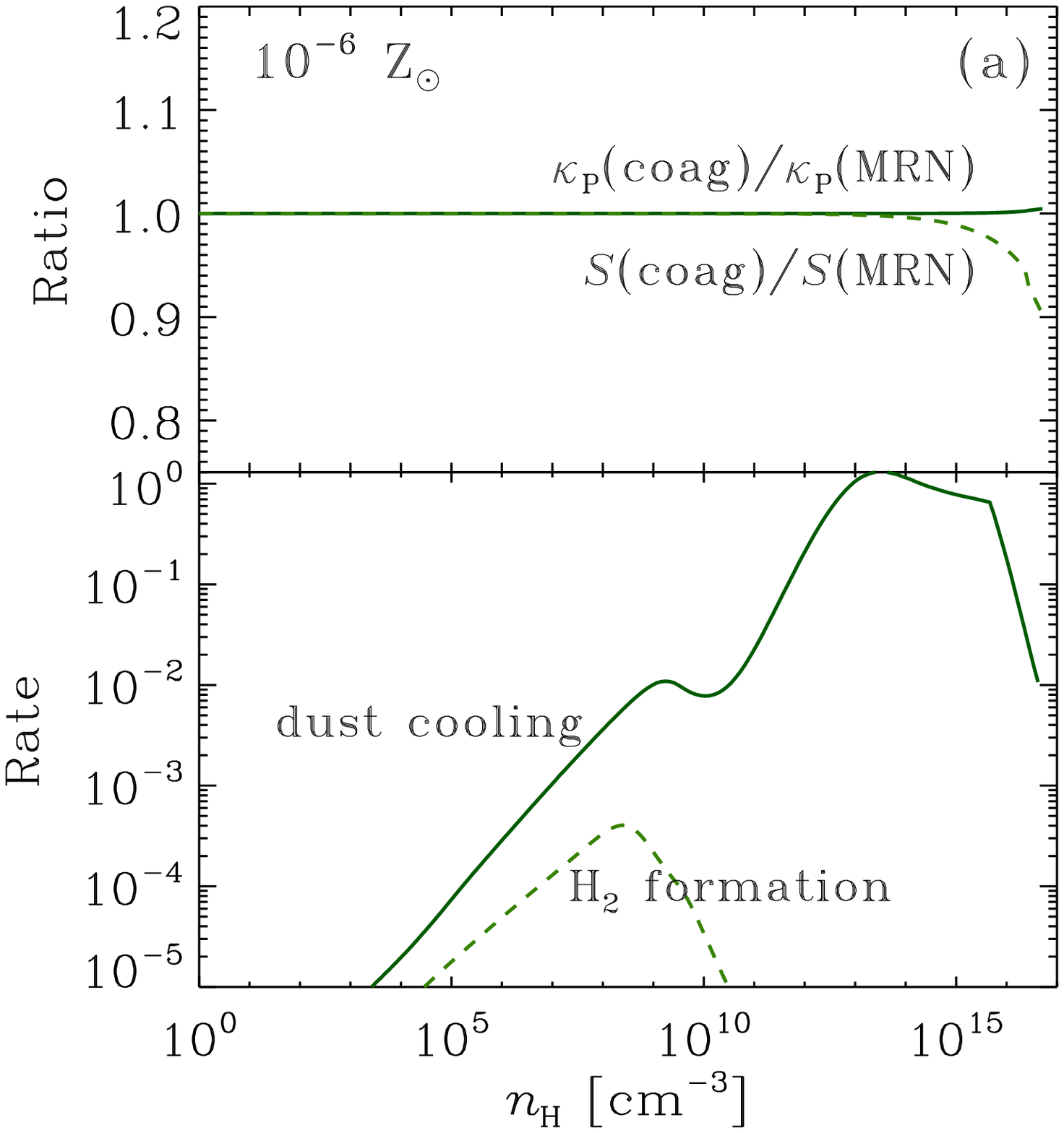}
\includegraphics[width=0.45\textwidth]{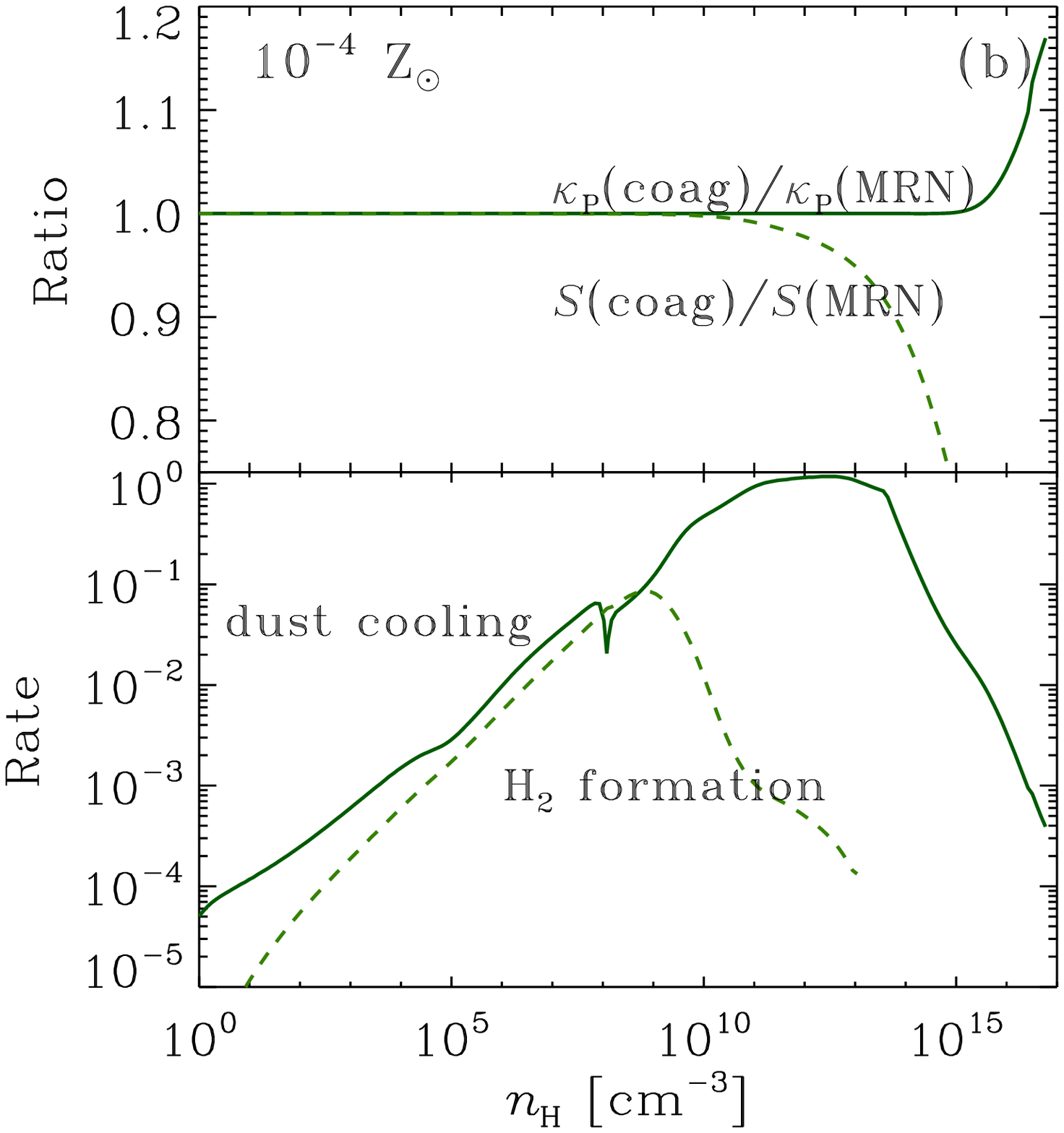}
\includegraphics[width=0.45\textwidth]{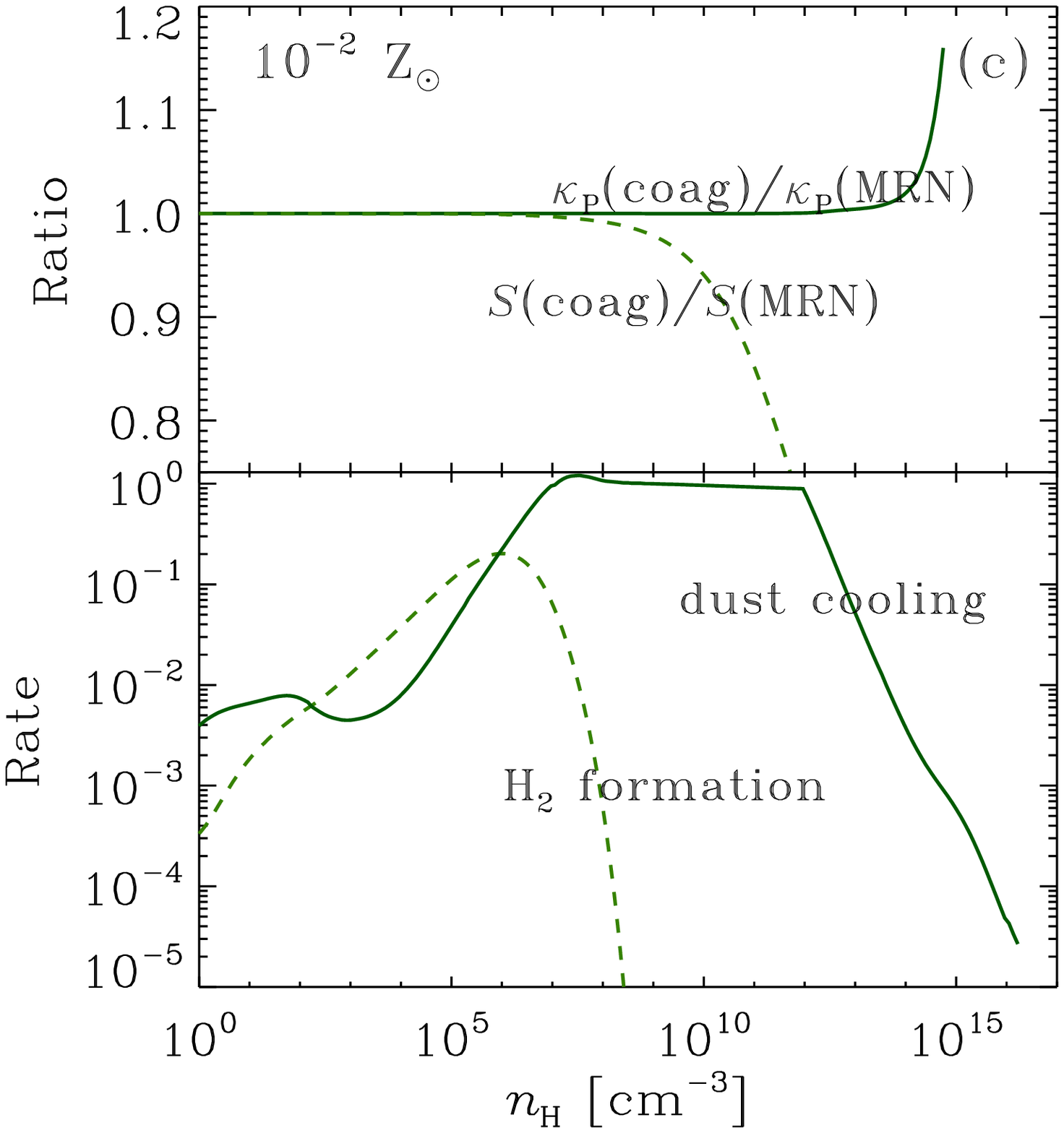}
\includegraphics[width=0.45\textwidth]{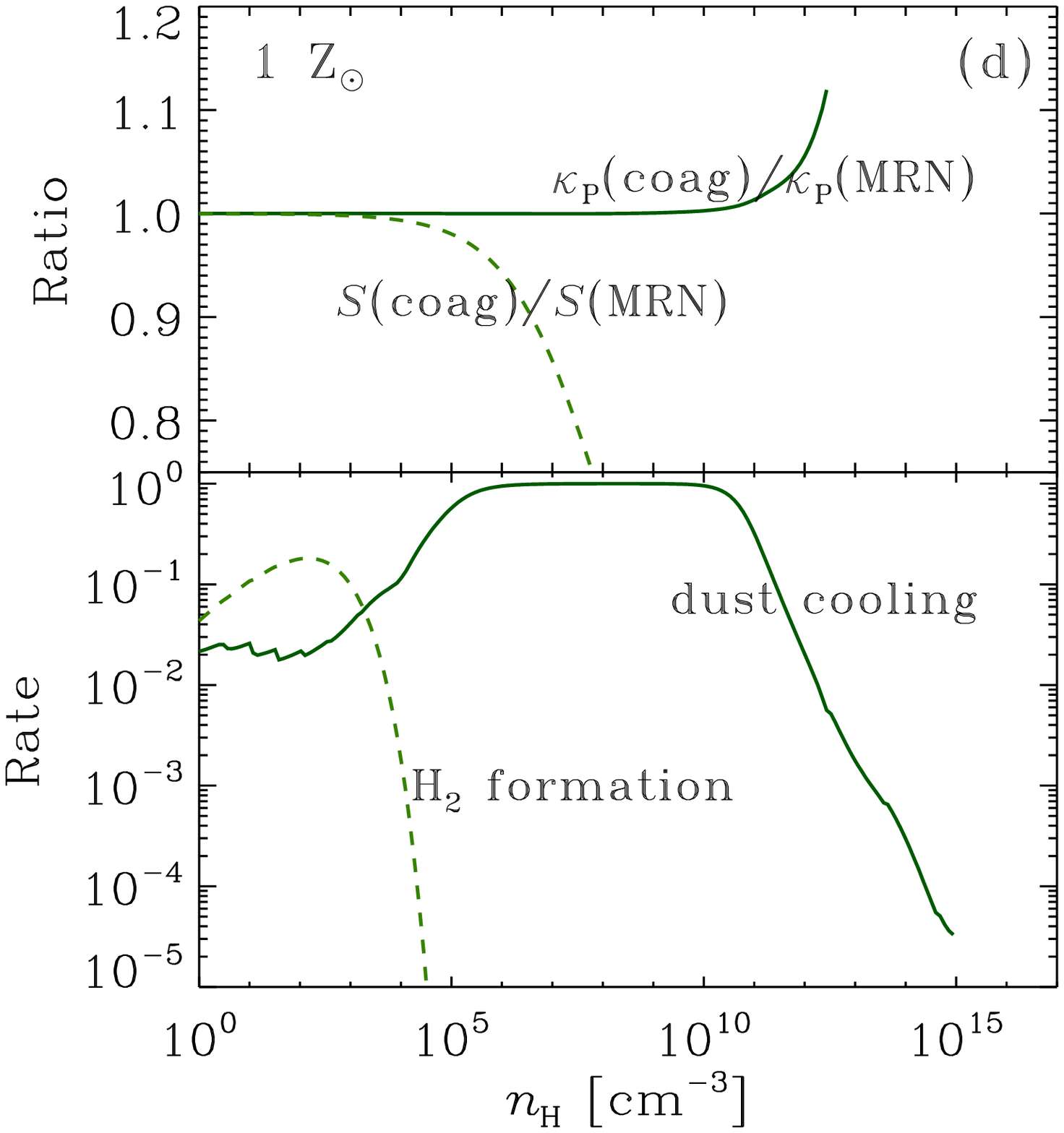}
 \caption{Upper panels: the evolutions of Planck mean 
dust opacity (solid line) and grain surface area
(dashed line). 
The values of these two quantities in the presence 
of coagulation are normalized to those without 
coagulation (i.e.\ with the initial MRN
grain size distribution preserved). Lower panels:
the ratio of dust cooling to
the total heating rate (solid line), and 
the increase rate of H$_2$ fraction by the
grain surface reaction in a free-fall time (dashed line). 
Each panel shows the result with metallicity
(a) $10^{-6}~\Zsun$, (b) $10^{-4}~\Zsun$,
(c) $10^{-2}~\Zsun$, and (d) 1 $\Zsun$.
}
 \label{fig:ratio}
\end{figure*}

Here we evaluate the effect of coagulation under the
density--temperature evolution calculated in the
previous section. Coagulation could affect the
thermal evolution of collapsing cores through the
changes of $S$ and $\kappap$. The former and
latter quantities have influence on the H$_2$
formation rate and the dust cooling rate,
respectively. In Fig.\ \ref{fig:ratio}, we show the
evolution of
these two quantities. Both quantities are divided by
the values without coagulation (i.e.\ the MRN grain
size distribution) and are denoted as
$\kappap (\mbox{coag})/\kappap (\mbox{MRN})$ and
$S(\mbox{coag})/S(\mbox{MRN})$.
The effects of coagulation appear
if these ratios deviate significantly from unity.
In fact, they change at high density, which indicates
that coagulation really takes place efficiently
enough to affect the grain opacity and the
grain surface area.

In Fig.\ \ref{fig:ratio}, we also show the dust
cooling rate relative to the total (compressional and chemical) 
heating rate and the increase rate of
H$_2$ fraction per free-fall time.
The latter is equal to
$(\mathrm{d}f_\mathrm{H_2}/d\ln\nH )_\mathrm{dust}$,
where $f_\mathrm{H_2}$ is the H$_2$ fraction 
($f_\mathrm{H_2}=1$ if hydrogen is
fully molecular), and the subscript ``dust''
indicates the contribution from grain surface
reaction.

\begin{figure*}
\includegraphics[width=0.45\textwidth]{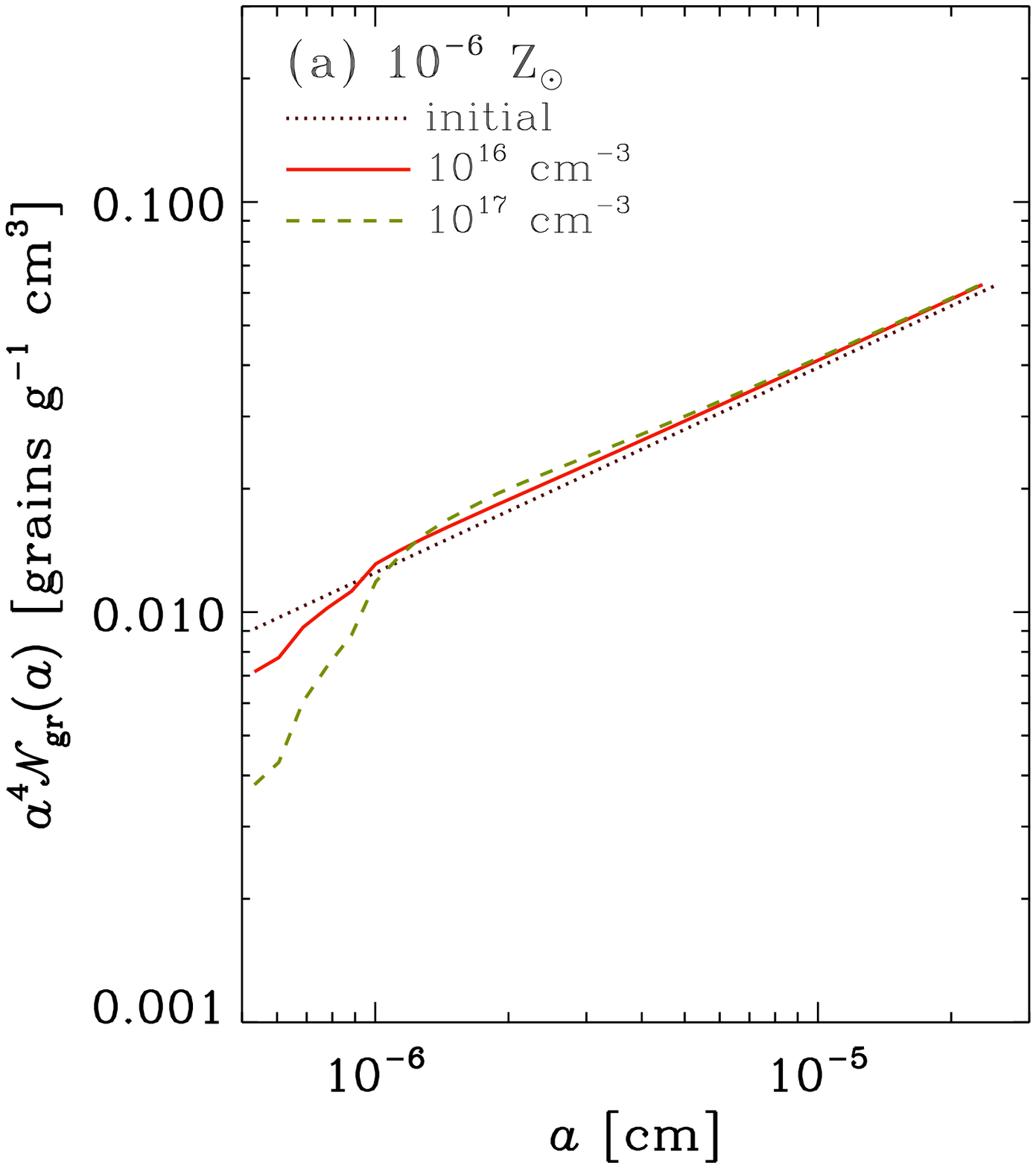}
\includegraphics[width=0.45\textwidth]{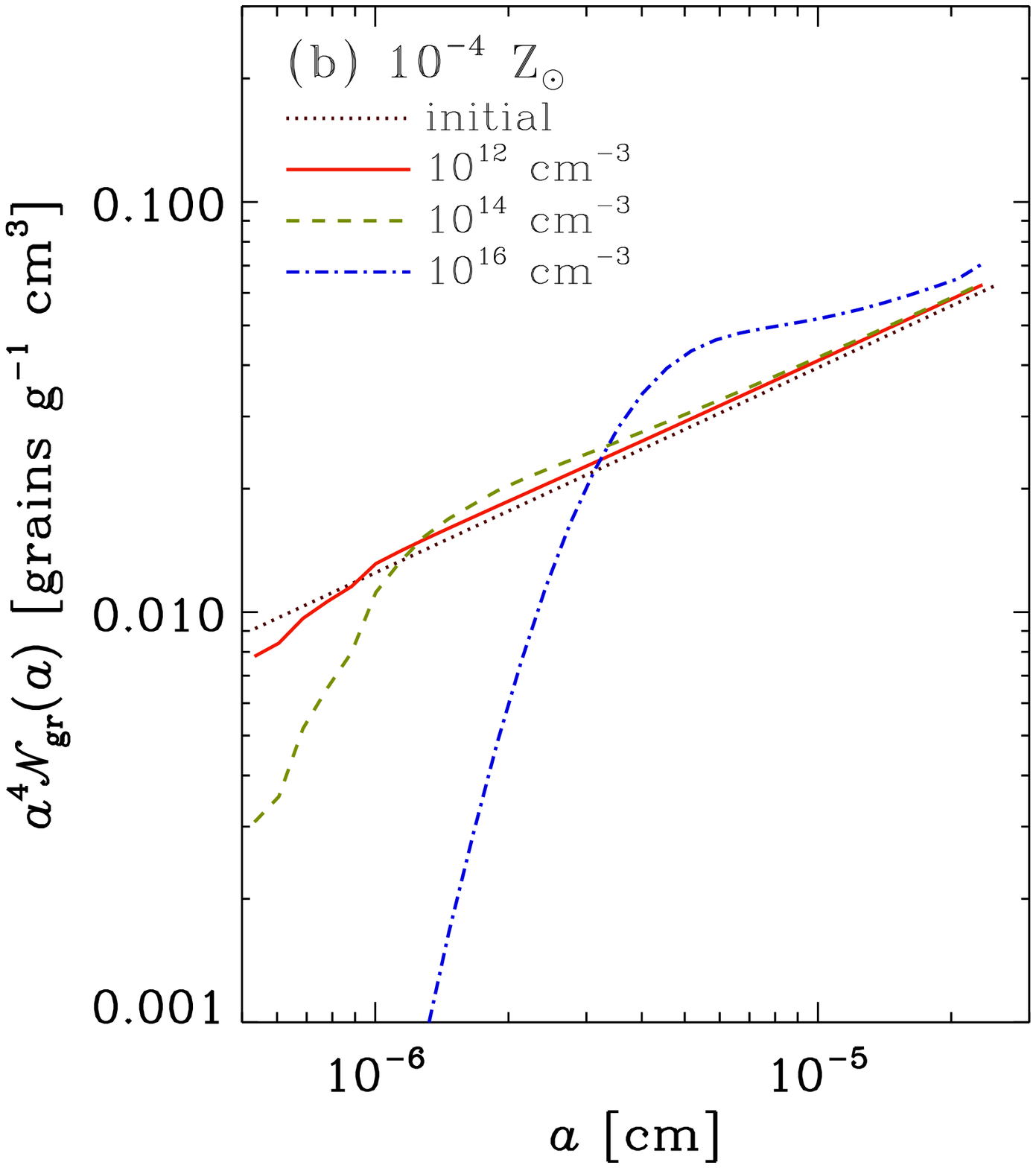}
\includegraphics[width=0.45\textwidth]{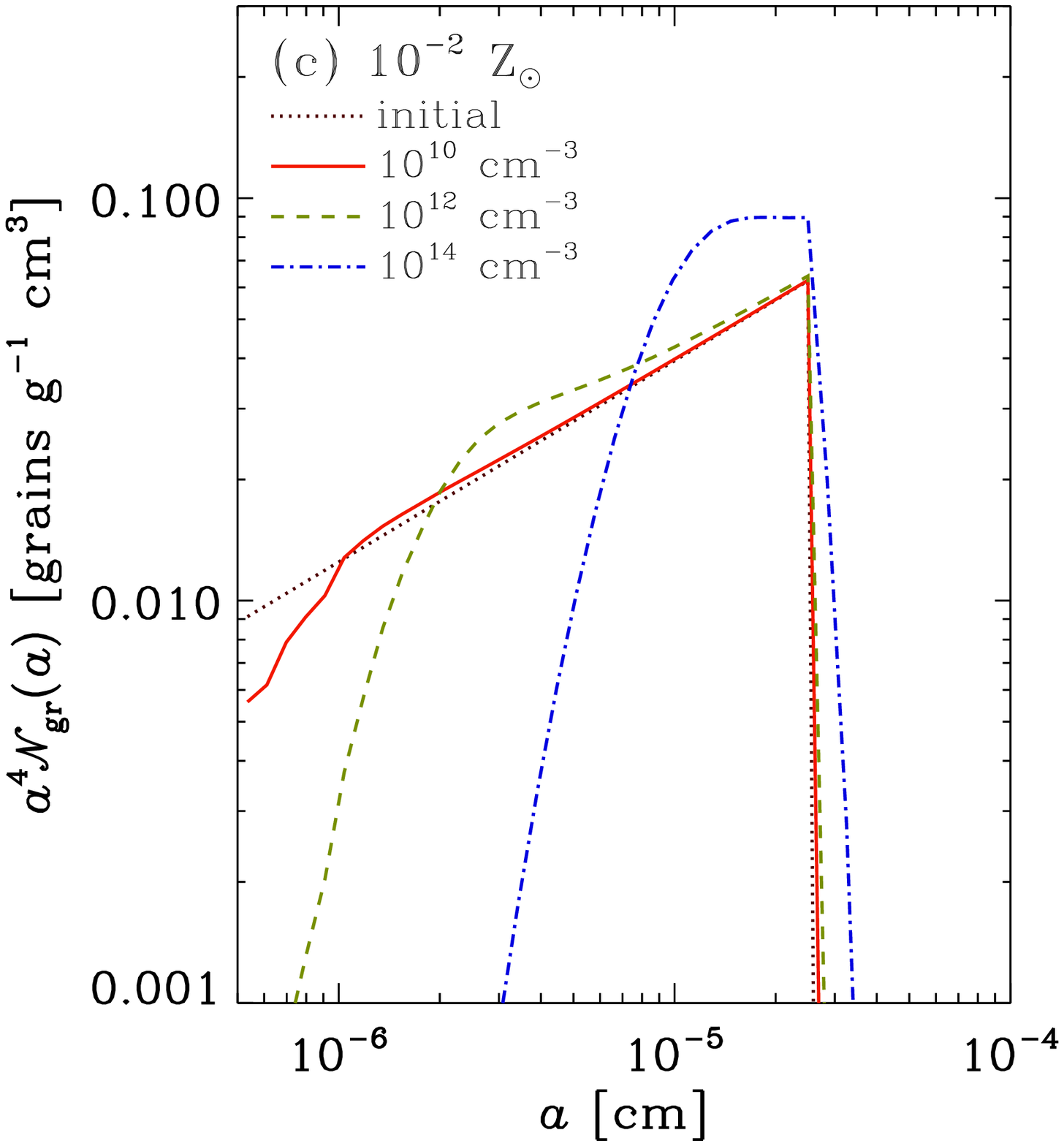}
\includegraphics[width=0.45\textwidth]{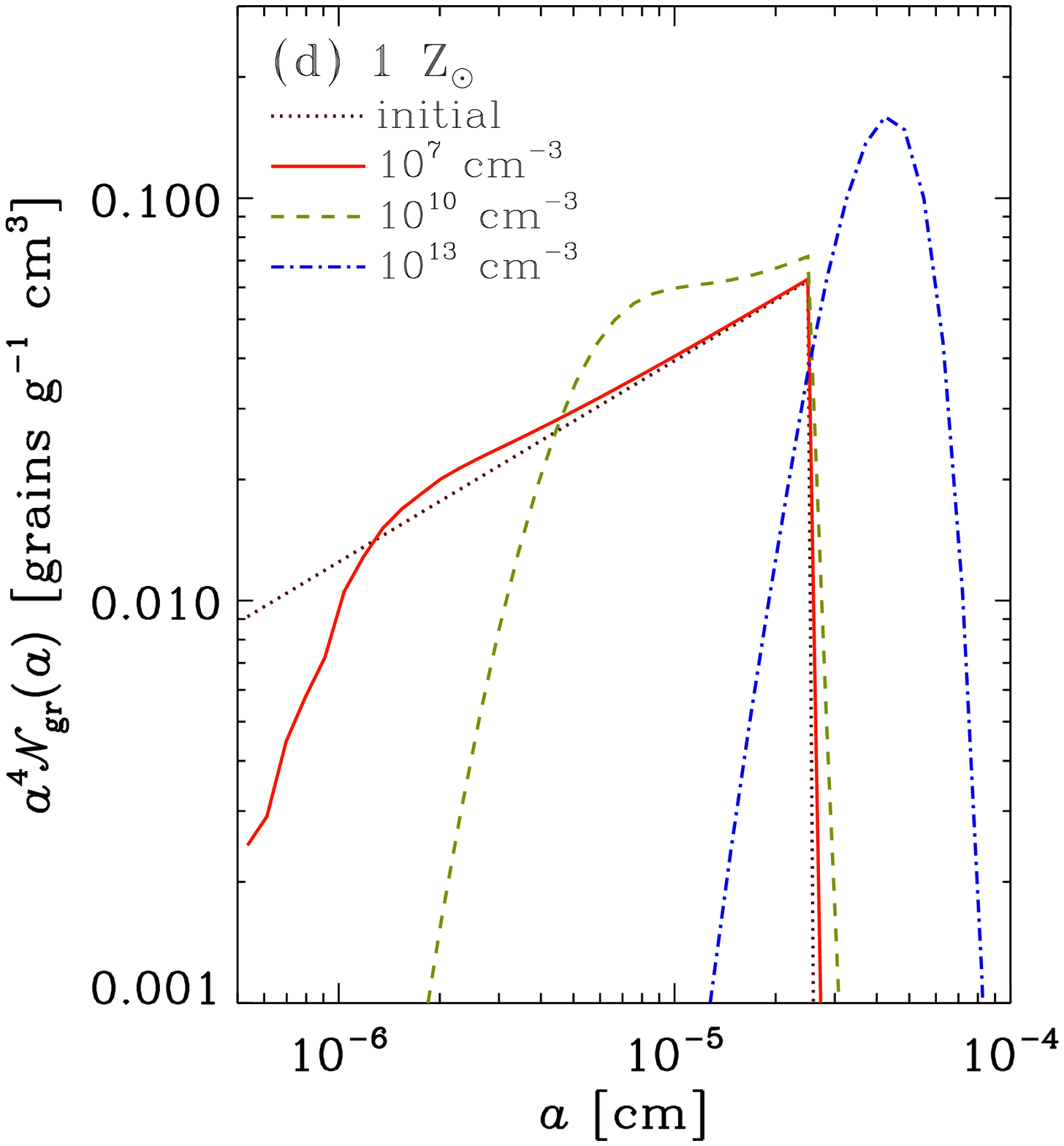}
 \caption{Evolutions of grain size distribution as
a function of hydrogen number density, which traces the
stage of gas collapse. The grain size distribution
function per unit dust mass, $\mathcal{N}_\mathrm{gr}(a)$, is
multiplied by $a^4$ to show the grain mass distribution
function. Panels (a), (b), (c), and (d)
show the results with $Z=10^{-6}$, $10^{-4}$, $10^{-2}$,
and 1 $\Zsun$, respectively. In each panel, the dotted
line presents the initial MRN size distribution, while
the solid, dashed, and dot-dashed lines represent
various evolutionary stages at the densities indicated in
each panel.
}
 \label{fig:size}
\end{figure*}

In order to have an idea about how coagulation
proceeds, we also show the grain size distributions
in Fig.\ \ref{fig:size}, where the grain size
distribution is multiplied by $a^4$ to show the grain
mass distribution per logarithmic size (the largest
contribution to the dust mass comes from the grain
size range where $a^4\mathcal{N}_\mathrm{gr}$
peaks). By comparing
Fig.\ \ref{fig:size} with Fig.\ \ref{fig:ratio}, we
observe that the decrease of grains
with $a\sim a_\mathrm{min}$ is indeed seen at the
density range where $S(\mbox{coag})/S(\mbox{MRN})$
decreases in Fig.\ \ref{fig:ratio}. The increase of
$\kappap (\mbox{coag})/\kappap (\mbox{MRN})$, on the
other hand, occurs when the grains are coagulated to
$a\ga 10^{-5}$ cm. In the following, we will
discuss further details.

\subsection{Grain surface area and coagulation criterion}
\label{subsec:surface}

We compare the dashed lines of the upper and lower
panels (i.e.\ grain surface area vs.\  
H$_2$ formation rate by grain surface reaction) in
Fig.\ \ref{fig:ratio}. We observe that the drop of
grain surface area
generally occurs well after the H$_2$ formation 
on grain surface is completed. This means that coagulation has no
influence on the thermal evolution of collapsing cores either 
through H$_2$ line cooling or H$_2$ formation heating.
The main reason why H$_2$ formation is faster than
grain coagulation is that the thermal velocity of H atoms
is much larger than that of grains.

The coagulation time-scale $t_\mathrm{coag}$ can
be estimated by the grain--grain collision time-scale.
The condition that coagulation takes place on
a free-fall time-scale ($t_\mathrm{coag}<t_\mathrm{ff}$)
is expressed as $n_\mathrm{H}>n_\mathrm{H,coag}$, where
$n_\mathrm{H,coag}$ is called ``critical density for
coagulation'' in this paper. The critical density for
coagulation is estimated as
\begin{eqnarray}
n_\mathrm{H,coag} & = & 1.6\times 10^7\,\beta^{-2}\left(
\frac{\langle a\rangle}{10^{-6}~\mathrm{cm}}\right)^{5}
\left(\frac{s}{3~\mathrm{g}~\mathrm{cm}^{-3}}\right)^{3}
\nonumber\\
& & \times\left(\frac{Z}{\Zsun}
\right)^{-2}\left(\frac{T}{100~\mathrm{K}}\right)^{-1}~
\mathrm{cm}^{-3},\label{EQ:DENS_CR}
\end{eqnarray}
where $s$ is the grain material density. This equation
is derived in the Appendix, where the grain size is
represented by a single value $\langle a\rangle$
for the simplicity of estimation. Indeed as shown in
Figs.\ \ref{fig:ratio} and \ref{fig:size},
if $\nH >n_\mathrm{H,coag}$, coagulation
significantly decreases the grain surface area,
and the grain size distributions at $a\la 10^{-6}$~cm
are indeed affected by coagulation. This
criterion can be used
generally, as long as a grain size distribution
similar to MRN is adopted and the grain velocity
is thermal.
Equation (\ref{EQ:DENS_CR}) can also be
solved for $a$ under a given hydrogen number density
to obtain a condition for the grain size
affected by coagulation.

As shown in equation (\ref{EQ:DENS_CR}), if we assume
a smaller sticking efficiency ($\beta<1$),
$n_\mathrm{H,coag}$ increases in proportion to
$\beta^{-2}$. Thus, the conclusion
that coagulation has no influence on the thermal
evolution of collapsing cores is strengthened if
we adopt $\beta <1$.

\subsection{Grain opacity}\label{subsec:dustcooling}

We compare the solid lines of the upper and lower
panels (i.e.\ Planck mean opacity vs.\ dust cooling)
in Fig.\ \ref{fig:ratio}. We find that the
change of
$\kappap (\mbox{coag})/\kappap (\mbox{MRN})$ appears
after the epoch in which dust cooling
dominates. In other words, before
coagulation affects the grain opacity,
the contribution from
dust cooling has already declined because the system
becomes optically thick to dust continuum.
This indicates that the change of grain
opacity by coagulation has no influence on the
thermal evolution. In particular,
$\kappap (\mbox{coag})/\kappap (\mbox{MRN})$
is kept to be unity even after
$S(\mbox{coag})/S(\mbox{MRN})$ drops significantly.
This is because
$\kappap$ does not depend on the grain radius
under a fixed total grain mass.
In fact, from equation (\ref{eq:kappa_nu}),
$\kappap\propto a^2Q_\nu N_\mathrm{gr}$
($N_\mathrm{gr}$ is the total grain number per unit
dust mass) with $a$
evaluated at a typical grain radius, while
$Q_\nu\propto a$ if $\lambda =c/\nu\gg a$
\citep{draine84} and $N_\mathrm{gr}\propto a^{-3}$
from the conservation of the total dust mass.

Since the grain opacity is proportional to the grain
mass, Fig.~\ref{fig:size} is useful to understand
which grain size dominates the grain opacity.
The largest grains dominate the grain mass in the
MRN grain size distribution, which means that the
grain opacity is also dominated by the largest grains.

In all the metallicities except for $Z=10^{-6}~\Zsun$,
we find that
$\kappap (\mbox{coag})/\kappap (\mbox{MRN})$
increases at a high density depending on metallicity.
At this density, the dust temperature, which
is almost equal to the gas temperature, is
$\sim 100$--1000 K, and the radiation spectrum
has a peak around wavelengths of
$\lambda\sim 3$--30 $\mu$m.
At this wavelength range,
$Q_\nu /a$ of graphite increases as a function of $a$
if $a\ga 0.1~\mu$m
because of the contribution from magnetic dipole
radiation \citep{draine84}. Moreover, the fraction of
grains with $a\ga 0.1~\mu$m is enhanced after
coagulation. These two effects (temperature increase
and grain growth) are the reason for 
the increase of
$\kappap (\mbox{coag})/\kappap (\mbox{MRN})$.



\section{Conclusion}\label{sec:conclusion}

By calculating the thermal evolution and the grain
coagulation in collapsing cores, we have shown that
coagulation does not affect their
thermal evolution for any metallicity
($0\le Z\la 1~\Zsun$) as long as the grain
velocities are governed by thermal motions.

Although the above statement is the most important
conclusion in this paper, there are two things that
are worth raising. First, coagulation really takes
place when the density is high enough. Coagulation is
important at $\nH\ga 10^7$ cm$^{-3}$ for $Z=\Zsun$
and the density range with
efficient coagulation shifts to higher densities
because of lower grain abundance (i.e.\ lower
grain collision rate). The condition that
$t_\mathrm{ff}>t_\mathrm{coag}$ is equivalent to
$\nH>{10^7}(Z/\Zsun )^{-2}(T/100~\mathrm{K})^{-1}~
\mathrm{cm}^{-3}$. This can be widely used as the
criterion that coagulation occurs in a free-fall
time as long as the grain size distribution
is MRN and the grain velocities are
dominated by the thermal motions. Second, $\kappap$
is unaffected by coagulation up to high
densities where the contribution from dust cooling
drops. The reason why $\kappap$
is insensitive to coagulation is that
$\kappap$ is almost independent of the grain
radius if
$a\ll \lambda$ ($\lambda$ is the wavelength at which
the radiation spectrum peaks) holds and
the dust temperature is $\la 100$ K.

\section*{Acknowledgments}
We thank the anonymous referee for useful and constructive
comments which improved this paper considerably.
We thank A. Ferrara for helpful suggestions.
This study is supported in part by the Grants-in-Aid by the Ministry
of Education,
Culture, and Science of Japan (19047004, 21684007; KO).


\newpage

\appendix

\section{Derivation of equation (\ref{EQ:DENS_CR})}

We provide an estimate of coagulation time-scale. The
coagulation time-scale, $t_\mathrm{coag}$, can be
estimated by the time-scale on which a grain encounter
another grain:
\begin{eqnarray}
t_\mathrm{coag}\simeq
\frac{1}{\beta n_\mathrm{gr}\pi\langle a
\rangle^2v_\mathrm{gr}},
\label{eq:coag}
\end{eqnarray}
where we have assumed that the grains have typical
representative radius $\langle a\rangle$ and
velocity $v_\mathrm{gr}$, and $n_\mathrm{gr}$
is the
grain number density. We explicitly write the sticking
probability $\beta$, which is assumed to be 1 in the
text. The grain velocity can be estimated
by the thermal velocity,
\begin{eqnarray}
v_\mathrm{gr}\simeq\left(
\frac{8k_\mathrm{B}T}{\pi m_\mathrm{gr}}\right)^{1/2},
\label{eq:vgr}
\end{eqnarray}
where the typical grain mass $m_\mathrm{gr}$ is
estimated to be
\begin{eqnarray}
m_\mathrm{gr}\simeq\frac{4\pi}{3}\langle a\rangle^3s.
\label{eq:mgr}
\end{eqnarray}
The grain number density, $n_\mathrm{gr}$ can be related
to the dust-to-gas ratio as
\begin{eqnarray}
\frac{4}{3}\pi\langle a\rangle^3sn_\mathrm{gr}
=n_\mathrm{H}m_\mathrm{H}(1+4y_\mathrm{He})\mathcal{D}.
\label{eq:ngr}
\end{eqnarray}
Using equations (\ref{eq:vgr}), (\ref{eq:mgr}), and
(\ref{eq:ngr}), equation (\ref{eq:coag}) is written as
\begin{eqnarray}
t_\mathrm{coag} & \simeq & \frac{2\sqrt{6}\pi}{9}
\frac{\langle a\rangle^{5/2}s^{3/2}}
{\beta n_\mathrm{H}m_\mathrm{H}(1+4y_\mathrm{He})
\mathcal{D}(k_\mathrm{B}T)^{1/2}}\nonumber\\
& \simeq &
\frac{5.7\times 10^{18}}{n_\mathrm{H}}\beta^{-1}\left(
\frac{\langle a\rangle}{10^{-6}~\mathrm{cm}}\right)^{5/2}
\left(\frac{s}{3~\mathrm{g}~\mathrm{cm}^{-3}}\right)^{3/2}
\nonumber\\
& & \times\left(\frac{\mathcal{D}}{6\times 10^{-3}}
\right)^{-1}\left(\frac{T}{100~\mathrm{K}}\right)^{-1/2}
~\mathrm{s}.
\end{eqnarray}

Now we examine if coagulation takes place within the
collapsing timescale, i.e.\ the free-fall timescale.
The free-fall time-scale is numerically estimated as
\begin{eqnarray}
t_\mathrm{ff}\simeq
\frac{1.4\times 10^{15}}{\sqrt{n_\mathrm{H}}}~\mathrm{s}.
\end{eqnarray}
Significant coagulation occurs if
$t_\mathrm{coag}<t_\mathrm{ff}$ is satisfied, which is
equivalent to the condition
$n_\mathrm{H}>n_\mathrm{H,coag}$. The critical density
for coagulation, $n_\mathrm{H,coag}$, is given in
equation (\ref{EQ:DENS_CR}),
where we have replaced $\mathcal{D}/6\times 10^{-3}$ with
$Z/\Zsun$ by assuming
that dust-to-gas ratio is proportional to metallicity.

\bsp

\label{lastpage}


\begin{thebibliography}{99}
\bibitem[\protect\citeauthoryear{Blum}{2000}]{blum00}
    Blum, J. 2000, Space Sci.\ Rev., 92, 265
\bibitem[\protect\citeauthoryear{Bohren \& Huffman}{1983}]{bohren83}
    Bohren, C. F., \& Huffman, D. R. 1983, Absorption and Scattering
    of Light by Small Particles, Wiley, New York
\bibitem[\protect\citeauthoryear{Bromm et al.}{2001}]{bromm01}
    Bromm, V., Ferrara, A., Coppi. P. S. \& Larson, R. B. 2001, 
MNRAS, 328, 969
\bibitem[\protect\citeauthoryear{Bromm \& Larson}{2004}]{bromm04}
    Bromm, V., \& Larson, R. B. 2004, ARA\&A 42, 79
\bibitem[\protect\citeauthoryear{Chokshi, Tielens, \& Hollenbach}{1993}]{chokshi93}
    Chokshi, A., Tielens, A. G. G. M., \& Hollenbach, D. 1993, ApJ, 497, 806
\bibitem[\protect\citeauthoryear{Draine \& Lee}{1984}]{draine84}
    Draine, B. T., \& Lee, H. M. 1984, ApJ, 285, 89
\bibitem[\protect\citeauthoryear{Dominik \& Tielens}{1997}]{dominik97}
    Dominik, C., \& Tielens, A. G. G. M. 1997, ApJ, 480, 647
\bibitem[\protect\citeauthoryear{Hirashita \& Yan}{2009}]{hirashita09}
    Hirashita, H., \& Yan, H. 2009, MNRAS, 394, 1061
\bibitem[\protect\citeauthoryear{Hollenbach \& McKee}{1979}]{hollenbach79}
    Hollenbach, D., \& McKee, C. F. 1979, ApJS, 41, 555
\bibitem[\protect\citeauthoryear{Jura}{1975}]{jura75}
    Jura, M. 1975, ApJ, 197, 575
\bibitem[\protect\citeauthoryear{Laor \& Draine}{1993}]{laor93}
    Laor, A., \& Draine, B. T. 1993, ApJ, 402, 441
\bibitem[\protect\citeauthoryear{Larson}{2005}]{larson05}
    Larson, R. B. 2005, MNRAS, 359, 211
\bibitem[\protect\citeauthoryear{Mathis, Rumpl, \& Nordsieck}{1977}]{mathis77}
    Mathis, J. S., Rumpl, W., \& Nordsieck, K. H. 1977, ApJ, 217, 425
    (MRN)
\bibitem[\protect\citeauthoryear{Meakin \& Donn}{1988}]{meakin88}
    Meakin, P., \& Donn, B. 1988, ApJ, 329, L39
\bibitem[\protect\citeauthoryear{Nozawa et~al.}{2003}]{nozawa03}
    Nozawa, T., Kozasa, T., Umeda, H., Maeda, K., \& Nomoto, K. 2003,
    ApJ, 598, 785
\bibitem[\protect\citeauthoryear{Omukai}{2000}]{omukai00}
    Omukai, K. 2000, ApJ, 534, 809
\bibitem[\protect\citeauthoryear{Omukai et~al.}{2005}]{omukai05}
    Omukai, K., Tsuribe, T., Schneider, R. \& Ferrara, A. 2005, ApJ, 626,
    627
\bibitem[\protect\citeauthoryear{Ormel et al.}{2009}]{ormel09}
    Ormel, C. W., Paszun, D., Dominik, C., \& Tielens, A. G. G. M. 2009,
    A\&A, in press
\bibitem[\protect\citeauthoryear{Ormel, Spaans, \& Tielens}{2007}]{ormel07}
    Ormel, C. W., Spaans, M., \& Tielens, A. G. G. M. 2007, A\&A, 461,
    215
\bibitem[\protect\citeauthoryear{Ossenkopf}{1993}]{ossenkopf93}
    Ossenkopf, V. 1993, A\&A, 280, 617
\bibitem[\protect\citeauthoryear{Schneider et al.}{2002}]{schneider02}
    Schneider, R., Ferrara, A., Natarajan, P. \& Omukai, K., 2002, ApJ,
    571, 30
\bibitem[\protect\citeauthoryear{Schneider, Ferrara, \& Salvaterra}{2004}]
{schneider04}
    Schneider, R., Ferrara, A., \& Salvaterra, R. 2004, MNRAS, 351, 1379
\bibitem[\protect\citeauthoryear{Schneider et al.}{2006}]{schneider06}
    Schneider, R., Omukai, K., Inoue, A. K., \& Ferrara, A. 2006, MNRAS,
    369, 1437
\bibitem[\protect\citeauthoryear{Spitzer}{1978}]{spitzer78}
    Spitzer, L., Jr 1978, Physical Processes in the Interstellar
    Medium, New York, Wiley
\bibitem[\protect\citeauthoryear{Todini \& Ferrara}{2001}]{todini01}
    Todini, P. \& Ferrara, A. 2001, MNRAS, 325, 726
\bibitem[\protect\citeauthoryear{Tielens \& Hollenbach}{1985}]{tielens85}
    Tielens, A. G. G. M. \& Hollenbach, D. J. 1985, ApJ, 291, 722
\bibitem[\protect\citeauthoryear{Weidenschilling \& Ruzmaikina}{1994}]{weidenschilling94}
    Weidenschilling, S. J., \& Ruzmaikina, T. V. 1994, ApJ, 430, 713
\bibitem[\protect\citeauthoryear{Yan, Lazarian, \& Draine}{Yan et al.}{2004}]{yan04}
    Yan, H., Lazarian, A., \& Draine, B. T. 2004, ApJ, 616, 895
\end{thebibliography}
\end{document}